\documentclass[]{article}
 
 \usepackage[utf8]{inputenc} 
 
 \usepackage{geometry} 
 \usepackage{graphicx}
 
 \usepackage{booktabs} 
 \usepackage{array} 
 \usepackage{paralist} 
 \usepackage{verbatim} 
 \usepackage{subfig} 
 
 \usepackage{fancyhdr} 
 \usepackage{sectsty}
 \allsectionsfont{\sffamily\mdseries\upshape} 
 
 \usepackage[nottoc,notlof,notlot]{tocbibind} 
 \usepackage[titles,subfigure]{tocloft} 
 
\author{ A. Sandoval-Villalbazo and A.R. Sagaceta-Mejía
	\bigskip \\
	Department of Physics and Mathematics,  U. Iberoamericana \\ M\'exico City, M\'exico.}

\title{On the existence of the Brillouin peaks in a simple dilute dissipative gas}

\begin{document}

\maketitle
\begin{abstract}
\noindent Light scattering due to the interaction of photons and acoustic waves present in a dilute inert gas is analyzed through the use of irreversible thermodynamics. The dispersion relation, which governs the dynamics of the density fluctuation of the gas allows the establishment of a simple criterion for the corresponding  Rayleigh-Brillouin spectrum to be observed. The criterion here proposed allows a clear physical interpretation and suggests generalizations for other interesting physical scenarios.
\end{abstract}

\section{Introduction}

\noindent Density fluctuations in a dilute gas in local equilibrium cause light scattering due to  Doppler interactions between the incoming photons and the acoustic modes of the fluid. This effect was first predicted by L. Brilouin \cite{Brillouin} and described theoretically by Landau and Placzek \cite{Landau1}. The spectrum reflects the dynamic behavior of density fluctuations and constitutes an experimental test of linear irreversible thermodynamics. 

It is well-known that, in the Euler regime, the transport equations that describe the fluctuations of the local thermodynamic variables of a simple static fluid read \cite{CC}:

\begin{equation}
	\frac{\partial}{\partial t}\left(\delta \rho\right)+\rho_{0}\left(\delta\theta\right)=0,
	\label{L1}
\end{equation}

\begin{equation}
	\frac{\partial}{\partial t}\left(\delta\theta\right)+  C^2_{T}
	\nabla^2 \left(
	\frac{\delta T}{T_{0}}
	\right)
	+ C^2_{T} \nabla^2 \left(\frac{\delta \rho}{\rho_{0}} \right)
	=0,
	\label{L2}
\end{equation}

\begin{equation}
\frac{\partial}{\partial t}
\left(\delta T\right)
+\frac{2}{3} T_{0} \left( \delta \theta\right) = 0,
\label{L3}
\end{equation}

where $\delta \rho$ corresponds to the density fluctuations around the equilibrium state $\rho_{0}$, $\delta T$ the temperature fluctuations around $T_{0}$ and $\delta \theta= \nabla \cdot \left(\delta \vec{u}\right) $ the expansion rate of the gas. These equations predict the existence of sound waves in the fluid which propagate with speed characteristic speed $C_T $, given by 
\begin{equation}
C_T=\sqrt{\frac{k T}{m}}.
\label{CT}
\end{equation}
In Eq. (\ref{CT}), $m$ is the individual mass of the particles present in the system, $T$ is the temperature and $k$ is the Boltzmann constant.

The interaction of light with the waves present in a dilute static fluid is rather difficult to measure \cite{Berne,Mountain}. For a given spatial  mode $k$, the specific fluid-photon interaction depends on the density of the fluid. In this context, the \textit{Brillouin scattering} is far more easy to be detected for high density systems rather than dilute gases. On the other hand, low density fluids are frequent in astrophysical scenarios in which structures may be formed at very long wavelengths \cite{Nos3}. This motivates the analysis of the conditions for the existence of acoustic waves in dilute fluids. 

In the presence of dissipation, Eqs.  (\ref{L1}-\ref{L3}) must include transport coefficients that take into account viscosity and heat conductivity, these effects may prevent the existence of sound waves in a dilute gas. 

The purpose of the present work is to establish a necessary  condition that must be satisfied in order to guarantee the existence of complex roots in the dispersion relation corresponding to the linearized transport system. If only real roots are present, no Brillouin doublet can be observed in a given experimental array. To accomplish this task the paper has been divided as follows: In section two, the dispersion relation that describes the dynamics of the fluctuations present in a simple dissipative fluid  is   expressed in terms of only one dissipative parameter, the relaxation time $\tau_r$. In section 3, the necessary condition for the existence of three real roots of the dispersion relation is established and a numerical example relevant in low density physics is presented. Final remarks are included in section 4.

\section{Dispersion relation in the presence of dissipation}
For a simple dilute gas consisting of hard spheres, Eq. (\ref{L1}) remains invariant in the presence of dissipation, while Eqs. (\ref{L2}-\ref{L3}) become  \cite{Berne}:

\begin{equation}
 \frac{\partial \left(\delta \theta\right)}{ \partial t} +C^2_T\left(\frac{\nabla^2 \left(\delta T\right)}{T_0}+ \frac{\nabla^2\left(\delta \rho\right)}{\rho_0}\right)=-\frac{1}{\rho_0} \nabla \cdot \left(\delta \Pi\right),
\label{Ld2}
\end{equation}

\begin{equation}
\frac{\partial \left(\delta T\right)}{ \partial t} +\frac{2}{3} T_0 \left(\delta \theta\right) +\frac{2 m}{3 k \rho_0} \nabla \cdot \vec{J}_{[Q]}  = 0.
\label{Ld3}
\end{equation}

In Eq. (\ref{Ld2} - \ref{Ld3}), $\delta \Pi$ corresponds to the stress tensor fluctuations and $\vec{J}_{[Q]}$ corresponds to the heat flux. If kinetic theory is applied within the BGK approximation \cite{BGK}, two constitutive equations can be established, namely
$$
\delta \Pi= -\eta_s \delta \sigma,
$$
and 
$$
\delta \vec{J}_{[Q]} = -\kappa_{th} \nabla \left(\delta T\right),
$$

\noindent 
where $\sigma$ corresponds to the traceless symmetric part of the velocity gradient,  $\eta_s $ is the shear viscosity and $\kappa_{th}$ the thermal conductivity of the gas.

\noindent The introduction of the constitutive equations in the set  (\ref{Ld2} - \ref{Ld3}) leads to 

\begin{equation}
\frac{\partial \left(\delta \theta\right)}{ \partial t} -C^2_{T} \left(\frac{\nabla^2 \left(\delta T\right)} {T_0}+\frac{\nabla^2\left( \delta \rho\right)} {\rho_0}\right)-D_v \nabla^2 \left(\delta \theta\right)=0,
\label{Lb2}
\end{equation}

\begin{equation}
\frac{\partial \left(\delta T\right)}{ \partial t}-\frac{2}{3} T_0\left( \delta \theta\right)+ D_{th} \nabla^2\left( \delta T\right)=0,
\label{Lb3}
\end{equation}
denoting $ \tau_r $ the relaxation time of the gas, the transport coefficients in Eqs. (\ref{Lb2}-\ref{Lb3}) become  $D_v=C^2_T \tau_r$  and $D_{th}=\frac{5}{3} C^2_T \tau_r $. Moreover, if the Fourier-Laplace transform of the fluctuation $\delta x$ is defined as:
 
$$
\delta \tilde{X}(\vec{q},s)=\int^{\infty}_{0} \int^{\infty}_{-\infty} \delta x(\vec{r},t) e^{i \vec{q} \cdot \vec{r}} e^{-s t} d\vec{r} dt,
$$
the system of transport equations can be  algebraically expressed as

\begin{equation}
A \cdot \delta \tilde{X}\left(\vec{q},s\right)=\delta \tilde{X}\left(\vec{q},0\right),
\label{Sys1}
\end{equation}
where
$$
\delta \tilde{X}=\left( \delta \tilde{ \rho}, \delta \tilde{ \theta},  \delta\tilde{T}\right),
$$
and

\begin{center}
	
\(A=\left(
\begin{array}{ccc}
s  & \rho_0   & 0 \\
-\frac{C_T^2 q^2}{\rho_{0}} & s+ C_T^2\tau_r  q^2 & - \frac{C_T^2q^2}{T_0}  \\
0 & \frac{2}{3} & \frac{s}{T_0}+\frac{5}{3}\frac{C_T^2\tau_r q^2}{T_0}  \\
\end{array}
\right)\).		
 
\end{center}

\noindent The system (\ref{Sys1}) governs the dynamics of the fluctuations of the local thermodynamic variables of the gas.  Real values for $s$ in the \textit{dispersion relation} $\det (A)=0$ are identified with exponentially decaying modes.  The resulting expression reads: 

\begin{equation}
 s^3+ \frac{8}{3} C^2_T \tau_r q^2 s^2+\left(\frac{5}{3} C^4_T q^4 \tau_r^2+\frac{5}{3} C^2_T q^2\right)s+\frac{5}{3}C^4_T \tau_r q^4 =0.
\label{Disp1}
\end{equation} 
In the next section, a simple geometrical analysis of the dispersion relation is applied in order to establish a necessary condition for the existence of two different complex roots, which in turn correspond to the presence of acoustic waves in the gas.

\section{Analysis of the dispersion relation}
Defining Eq. (\ref{Disp1}) as a function of $s$, the resulting expression reads
$$
f(s)= s^3+ \frac{8}{3} C^2_T \tau_r q^2 s^2+\frac{5}{3} \left(C^4_T q^4 \tau_r^2+ C^2_T q^2\right)s+\frac{5}{3}C^4_T \tau_r q^4 ,
\label{Disp3}
$$
the first two derivatives of $f(s)$ read
\begin{equation}
f'(s)=3s^2+ \frac{16}{3} C^2_T q^2 \tau_r s+ \frac{5}{3} C^4_T q^4 \tau_r^2+\frac{5}{3} C^2_T q^2, \qquad 
f''(s)=6 s+\frac{16}{3} C^2_T q^2 \tau_r.
\end{equation}
The inflection point of $f(s)$ is always located at $s=-\frac{8}{9} C^2_T q^2 \tau_r \simeq  -C^2_T q^2 \tau_r$.  In the absence of dissipation, $f(s)$ is symmetric with respect of the inflection point, which in this case  is located at the origin.
A necessary condition for the existence of three different roots for $f(s)$ is
\begin{equation}
 \Delta=-45 q^2 C^2_T+19  C^4_T q^4 \tau_r^2 >0,
\end{equation}
or 
\begin{equation}
q^2 >\frac{45}{19 C_T^2  \tau^2_r },
\end{equation}
In the case of astrophysical systems such as globular clusters, densities are quite low and the temperatures are well beyond the ionization values. In this kind of systems acoustic waves may appear if the wavenumber $q$ is low enough. 

\begin{figure}[h!]
	\centering		
	\includegraphics[height=7cm]{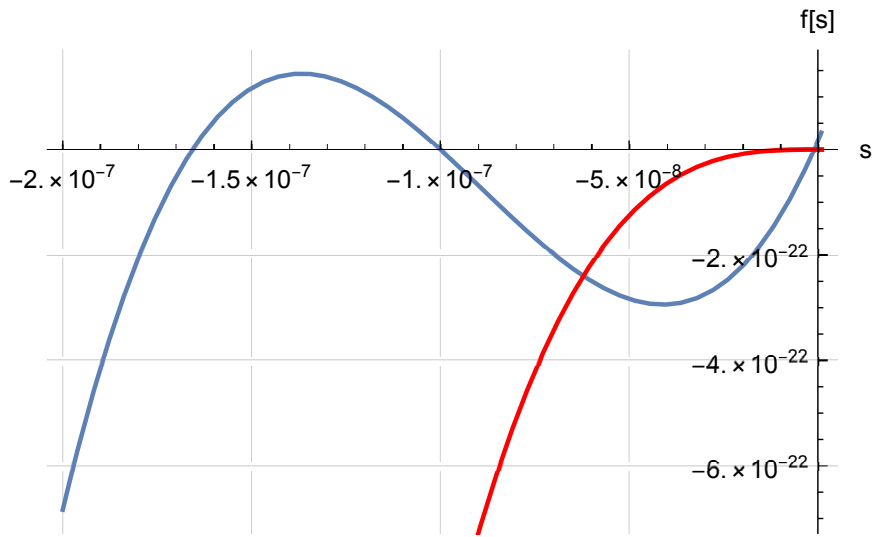}
	\caption{Dispertion relation  for $C_T=10^2$ m/seg and $\tau=10^9$ seg. The blue curve corresponds to  $q=10^{-10}$1/m (three real negative roots, no acoustic waves) and the  red curve to $q=10^{-13}$1/m (one real negative root and acoustic waves),}
	\label{fig:1}
\end{figure}

In Figure \ref{fig:1}, the blue line corresponds to the dispersion relation (Eq. (\ref{Disp3})) for $q=10^{-10} \frac{1}{m} $, $C_T=10^{2}\frac{m}{seg}$ and $\tau_r=10^9 seg$. In this case $C_T q \tau_r=10 $ and no acoustic waves are present for this mode. The red curve corresponds to the dispersion relation for the same values of $C_T$ and $\tau_r$, but with $q=10^{-13} \frac{1}{m}$. In this case two complex roots appear, corresponding to acoustic waves of very large wavelengths.

\section{Final Remarks}
It is very hard to find in the literature simple examples in which a discussion of the existence of the roots precedes to the pursue of the solutions of the dispersion equation for dilute mixtures. If dissipation is strong enough, it can prevent the formation of acoustic waves. In fact, the ideas contained in this paper lead to the establishment of a cut-off wavelength that depends on the isothermal speed of sound in the gas. 

The extension of this criterion in the case of a single self-gravitating fluid can be analyzed through the dispersion relation \cite{Nos1} :
\begin{equation}
f(s)=s^3+(D_v+D_{th}) q^2 s^2-\left(\frac{5}{3}C^2_T q^2- D_v D_{th} q^4-4 \pi G \rho_0\right) s- 4 \pi G \rho_0 D_{th}q^2+\frac {5}{3}D_{t}C^2_Tq^4,
\end{equation}
and its first two derivatives:
\begin{eqnarray}
f'(s)=3s^2+2 (D_v+D_{th}) q^2 s-\left(\frac{5}{3}C^2_T q^2- D_v D_{th} q^4-4 \pi G \rho_0\right),\\
f''(s)=6s+2(D_v+D_{th}) q^2 .
\end{eqnarray}
It is interesting to notice that the \textit{necessary and sufficient condition} for the existence of three different real roots reduces to the inequality:

$$
f(s^+) f(s^{-})<0,
$$
where
\begin{equation}
s^{\pm}=-\frac{(D_v+D_{th}) q^2}{3}\pm  \frac{1}{3}\sqrt{5 C_T^2q^2+(D_v+D_{th})^2q^4-3(D_v D_{th}q^4+4\pi G\rho_0) }.
\label{eq:last}
\end{equation}
The simplified expression for the discriminant of eq. (\ref{eq:last}) is given by
\begin{equation}
b(q)=\frac{5}{3} C^2_{T} q^2\left(1+C^2_{T} q^2 \tau^2_{r}\right)-4  \pi G \rho<0.
\end{equation}
Taking the value of the Jeans wavenumber as $\frac{4 \pi G \rho_0}{C^2_T}$ \cite{Nos2}, it is easily noticeable that 
\begin{equation}
b(q_J)\simeq-48\pi G \rho_0,
\end{equation}
and no unstable modes appear at the ordinary critical wavelength. 

We consider that this algebraic approach to the analysis of the Brillouin peaks for dilute fluids is promising and useful for students and researchers interested the subject.

\section{Acknowledgments}
This work has been supported by the Applied Research Institute of Technology
(INIAT) of U. Iberoamericana, Mexico.
\nocite*{}
\bibliographystyle{ieeetr} 
\bibliography{Alf2021}

\end{document}